\documentclass[%
 reprint,
 amsmath,amssymb,
 aps,
prl,
]{revtex4-2}

\usepackage{graphicx}
\usepackage{dcolumn}
\usepackage{bm}
\usepackage{color}

\newcommand{\imag}{\mathrm{i}}
\newcommand{\diff}{\mathrm{d}}

\begin{document}

\preprint{APS/123-QED}

\title{Turbulent Thermal Equilibration of Collisionless Magnetospheric Plasmas}
\author{Ryusuke Numata}
\email{ryusuke.numata@gmail.com}%
\affiliation{%
Graduate School of Information Science, University of Hyogo, Kobe 650-0047, Japan
}%

\date{\today}

\begin{abstract}
How thermal equilibrium is determined in a weakly collisional plasma is a fundamental question in plasma physics. This letter shows that the turbulence driven by the magnetic curvature and density gradient tends to equilibrate the temperature between species without collisions in a magnetospheric plasma.
The classical stability analysis in terms of energetic consideration reveals the interchangeable roles of electrons and ions for destabilization depending on their temperatures. Nonlinear gyrokinetic simulations confirm that the higher-temperature destabilizing species gives free energy to heat the other species to achieve the equal temperature state.
\end{abstract}

\maketitle


A planetary magnetosphere is a peculiar plasma environment: A high-temperature plasma is confined in a strongly inhomogeneous dipolar magnetic field generated by a planet.
Inspired by nature, the dipole confinement concept as a fusion device was proposed~\cite{HasegawaChenMauel_90} and later levitated ring dipole confinement devices have been built and operated for years~\cite{BoxerEllsworthGarnier_10, YoshidaSaitohMorikawa_10}.
Currently, it is ready to study this natural complex system in detail by in-situ spacecraft observations and comprehensive diagnostics of experimental devices
to gain insight into the interplay between the inhomogeneous field and plasma dynamics.
In this letter, we consider a self-organized state of the magnetospheric plasma in terms of energetics. 
It is shown that the favorable state is a thermally equilibrated state by turbulent energy exchange between species.

A prominent phenomenon induced by the inhomogeneity of magnetic configuration is the inward pinch whereby particles are transported against the density gradient, formally violating Fick's law. The gyrokinetic theory was applied to analyze the linear stability and transport in the magnetospheric plasmas and revealed that the electrostatic entropy mode driven by the magnetic curvature and density gradient is responsible for turbulent transport~\cite{SimakovCattoHastie_01, *SimakovHastieCatto_02}. Later, gyrokinetic simulations of the entropy mode in the dipole configuration successfully demonstrated the existence of a particle pinch regime~\cite{KobayashiRogersDorland_09, *KobayashiRogersDorland_10}. 
An interesting explanation of this phenomenon in terms of the statistical mechanical point of view of Hamiltonian systems was proposed~\cite{SatoYoshida_16}. The {\it up-hill diffusion} results from entropy minimization in properly constructed constrained phase space incorporating the magnetic field structure.

Turbulence also heats plasmas (e.g., ~\cite{Candy_13, BarnesAbiusoDorland_18}). The turbulent heating can be a possible candidate to determine a thermal equilibrium state in collisionless environments, such as the magnetosphere.
As will be shown later, the gyrokinetic formulation of the entropy mode in magnetospheric plasmas shows that electrons and ions behave in a quite similar manner independent of the mass ratio except for finite Larmor radius (FLR) effects. The role played by electrons and ions can be interchangeable, likewise the heating direction. We address which of the species is responsible for driving the system unstable and causing internal energy exchange.
The argument is essentially the same as~\cite{BarnesAbiusoDorland_18} for the universal instability where the instability is caused by wave-particle interactions along the field line and therefore the thermal speed rather than the temperature itself matters.

In the following, we first perform a linear analysis to determine the instability mechanism. Then, we examine the relation between the linear instability mechanism and the nonlinear heating of plasmas by performing nonlinear simulations.

We consider the Z pinch configuration as a prototype of magnetospheric plasmas. In the Z pinch, a plasma in a cylinder is confined by the azimuthal magnetic field $\bm{B}=B_{\theta}(r)\nabla \theta$. We take a circular magnetic field of the Z pinch with the curvature radius given by $R_{\mathrm{c}}$ and solve a local electrostatic gyrokinetic model for electrons and single-species ions. Since there is no variation along the field line, it is sufficient to consider a two-dimensional slab perpendicular to the magnetic field. Hereafter, we take the Cartesian coordinate where the $z$ is the direction along the magnetic field and $x$ is the direction of the background gradient, $\bm{B}=B_{0}(x) \nabla z$ and $(\nabla B_{0})/B_{0} = - (\diff \ln B_{0}/\diff x)\nabla x$~\footnote{In a planetary magnetosphere, $x$ points outward from the planet, and $y$, $z$, respectively, point eastward, northward.}. The gyrokinetic equation of the species $s=\mathrm{i}, \mathrm{e}$ for the non-Boltzmann part of the perturbed distribution function $h_s$ where $\delta f_s = - (q_s \phi/T_{0s}) f_{0s} + h_{s}$ is written as
\begin{multline}
\frac{\partial h_{s}}{\partial t} + (\bm{v}_{\mathrm{D},s}+\bm{v}_{\bm{E},s}) \cdot \nabla h_{s} \\ = 
\left(
\frac{\partial \langle \phi \rangle_{\bm{R}_{s}}}{\partial t}
+ \bm{v}_{\ast,s}^{\mathrm{T}} \cdot \nabla \left\langle \phi \right\rangle_{\bm{R}_s} \right)
\frac{q_{s} f_{0s}}{T_{0s}} +
C(h_{s}),
\end{multline}
complemented by the quasi-neutrality condition $\sum_{s} q_{s} \delta n_{s} = 0$ ($\delta n_{s}=\int \delta f_{s} \diff \bm{v}$ is the density perturbation). The notation is standard (See, e.g., \cite{NumataHowesTatsuno_10}). We assume the collision term $C$, including the artificial hyper-diffusion, is small. It is ignored in most of the following analyses, but employed only in nonlinear simulations just to achieve steady states.
We decompose variables in Fourier series, $h_{s}(\bm{R}_{s},v_{\parallel},v_{\perp},t) = \sum_{\bm{k}} h_{\bm{k},s}(v_{\parallel},v_{\perp}) e^{-\imag(\omega(\bm{k}) t - \bm{k}\cdot\bm{R}_{s})}$, $\phi(\bm{r},t)= \sum_{\bm{k}} \phi_{\bm{k}} e^{-\imag(\omega(\bm{k}) t - \bm{k}\cdot\bm{r})}$, where $\bm{r}=\bm{R}_{s} - \bm{v}\times\nabla z/\Omega_{\mathrm{c},s}$. The collisionless linear solution is given by
\begin{align}
h_{\bm{k},s} = \frac{\omega -\omega_{\ast,s}^{\mathrm{T}}}{\omega - \omega_{\mathrm{D},s}} J_{0}(\alpha_{s}) \frac{q_{s} \phi_{\bm{k}}}{T_{0s}} f_{0s}.
\label{eq:linear_soln}
\end{align}
The drift frequencies are defined by
\begin{align}
\omega_{\ast,s}^{\mathrm{T}} & = \bm{k}\cdot\bm{v}_{\ast,s}^{\mathrm{T}} = \omega_{\ast,n,s} (1 + \eta_{s} (v^2/v_{\mathrm{th},s}^2-3/2)), \\
\omega_{\mathrm{D},s} & = \bm{k} \cdot\bm{v}_{\mathrm{D},s} = \omega_{\kappa,s} v_{\parallel}^{2}/v_{\mathrm{th},s}^{2} + \omega_{\nabla B,s} v_{\perp}^{2}/(2 v_{\mathrm{th},s}^{2}),
\end{align}
with the velocity-independent drift frequencies are, respectively, given by
\begin{align}
\omega_{\ast,n,s} & = -\frac{1}{2} k_{y} \rho_{s} v_{\mathrm{th},s}/L_{n}, \\
\omega_{\kappa,s} & = -k_{y} \rho_{s} v_{\mathrm{th},s}/R_{\mathrm{c}}, \\
\omega_{\nabla B,s} & = -k_{y} \rho_{s} v_{\mathrm{th},s}/L_{B}.
\end{align}
The Larmor radius $\rho_{s}=v_{\mathrm{th},s}/\Omega_{\mathrm{c},s}$, the thermal velocity $v_{\mathrm{th},s}=\sqrt{2T_{0s}/m_{s}}$ and the cyclotron frequency $\Omega_{\mathrm{c},s}=q_{s} B_{0}/m_{s}$. Note that the cyclotron frequency is defined including the sign of charge. The electron drift frequencies are positive, indicating electrons drift in the positive $y$ direction. The background density and magnetic field gradient scales are $L_{n}\equiv-\diff \ln n_{0}/\diff x$ and $L_{B}\equiv-\diff \ln B_{0}/\diff x$ and the magnetic field curvature radius is $R_{\mathrm{c}}$. For the low-beta electrostatic case, the MHD equilibrium condition demands $L_{B}=R_{\mathrm{c}}$. In this case, we write $\omega_{\kappa,s}=\omega_{\nabla B,s}\equiv\omega_{B,s}$. The ratio of temperature and density gradients is $\eta_{s}=\diff \ln T_{0s}/\diff \ln n_{0}$. In the following analyses, we ignore $\eta_{s}$ for simplicity. The argument of Bessel function $J_{0}$ is $\alpha_{s}=k v_{\perp}/\Omega_{\mathrm{c},s}$ with $k=|\bm{k}|=\sqrt{k_{x}^{2}+k_{y}^{2}}$.

We introduce the standard notation of electrostatics in a dielectric medium. The polarization is defined by $q_{s} \delta n_{s} = -\nabla \cdot \bm{P}_{s}$ where $\bm{P}_{s}=\epsilon_{0} \chi_{s} \bm{E}$ using the electric susceptibility $\chi_{s}$ ($\epsilon_{0}$ is the vacuum permittivity) with $\bm{E}=-\nabla \phi$ being the electrostatic field. Since the polarization is time-dependent, it carries the polarization current $\bm{j}_{\mathrm{pol},s}=\partial \bm{P}_{s}/\partial t = \sigma_{s} \bm{E}$. The associated conductivity is given by $\sigma_{s} = - \imag \omega \epsilon_{0} \chi_{s}$. Using the solution of the gyrokinetic equation we obtain the susceptibility as
\begin{align}
\chi_{s} & = \frac{1}{k^{2} \lambda_{\mathrm{D},s}^{2}}
    \left(1 - \frac{\omega_{\ast,n,s} - \omega}{\omega_{B,s}} Q(\omega/\omega_{B,s},k\rho_{s}) \right),
\end{align}
where $\lambda_{\mathrm{D},s}\equiv\sqrt{\epsilon_{0}T_{0s}/(q_{s}^{2}n_{0})}$ is the Debye length and the integral $Q$ is given by
\begin{multline}
Q(\omega/\omega_{B,s}, k \rho_s) \\
= \frac{2}{\sqrt{\pi}} \int \frac{e^{-(\hat{v}_{\parallel}^2 + \hat{v}_{\perp}^{2})}}{\hat{v}_{\parallel}^2 + \hat{v}_{\perp}^2/2- \omega/\omega_{B,s}} J_{0}^{2} (k \rho_{s} \hat{v}_{\perp}) \hat{v}_{\perp} \diff \hat{v}_{\perp} \diff \hat{v}_{\parallel}.
\end{multline}
From the quasi-neutrality, which is the small $k\lambda_{\mathrm{D},s}$ limit of Gauss's law, the dispersion relation is given by $\chi \equiv \chi_{\mathrm{e}} + \chi_{\mathrm{i}} = 0$.
The integral $Q$ can be analytically evaluated if $k \rho_s \ll 1$~\cite{RicciRogersDorland_06}. If we ignore the FLR correction, $Q$ is simply equal to $-\Xi(\imag \sqrt{-\omega/\omega_{B,s}})^2$ where $\Xi(z)=\imag \sqrt{\pi} e^{-z^2} \mathrm{erfc}(-\imag z)$ is the plasma dispersion function, $\mathrm{erfc}(z)$ is the complementary error function.

The total energy of the system per unit volume $W$ is given by the sum of the energy of each species $W_{s}$, $W=\sum_{s}W_{s}$, where 
\begin{align}
W_{s} = \frac{1}{V} \int \frac{T_{0s} \delta f_{s}^{2}}{2f_{0s}} \diff \bm{v} \diff \bm{r}
    = \sum_{\bm{k}} \int \frac{T_{0s}}{2f_{0s}} \left| \delta f_{\bm{k},s} \right|^2 \diff \bm{v},
    \label{eq:particle_energy}
\end{align}
with $V=L_{x}L_{y}$ and $L_{x}$, $L_{y}$ being the box sizes in the $x$, $y$ directions.
The time evolution of the particle energy is given by
\begin{align}
\frac{\diff W_{s}}{\diff t} & = H_{s} + I_{s} - D_{s}, \\
H_{s} & = \frac{1}{V}\int q_{s} h_{s} \frac{\partial \langle \phi \rangle_{\bm{R}_{s}}}{\partial t} \diff \bm{v} \diff \bm{r}, \label{eq:heating} \\
I_{s} & = \frac{T_{0s}}{L_{n}} \varGamma_{s}
 = \frac{T_{0s}}{L_{n}} \frac{1}{V} \int h_{s} \left(- \frac{1}{B_{0}} \frac{\partial \langle \phi \rangle_{\bm{R}_{s}}}{\partial y}\right) \diff \bm{v} \diff \bm{r}, \\
D_{s} & = - \frac{1}{V} \int \frac{T_{0s} h_{s}}{f_{0s}} C(h) \diff \bm{v} \diff \bm{r},
\end{align}
where $H_{s}$ is the energy exchange between species (or the work done on the species $s$), $I_{s}$ is the energy injection, $\varGamma_{s}$ is the particle flux and $D_{s}$ is the collisional dissipation. We can show that $\sum_{s} H_{s} = 0$. Therefore, in the nonlinearly saturated state, $\sum_{s}(I_{s}-D_{s})=0$. The energy exchange term is known as the turbulent heating as it contributes to the increase of background temperature in a longer time scale~\cite{HowesCowleyDorland_06, Candy_13, BarnesAbiusoDorland_18},
\begin{align}
\frac{3}{2} n_{0} \frac{\diff T_{0s}}{\diff t'} + \frac{\partial \overline{\mathcal Q}_{s}}{\partial x'} = \overline{H_{s}},
\end{align}
where $t'$, $x'$ denote the macro-scale coordinates compared with the microscopic gyrokinetic $t$, $x$, the overline denotes a long-time average in $t$ and ${\mathcal Q}_{s}$ is the turbulent heat flux.

Substituting the collisionless linear solution \eqref{eq:linear_soln} into \eqref{eq:particle_energy}, we obtain 
\begin{align}
2 \Im(\omega) W_{\bm{k},s} = \frac{\epsilon_{0}}{4} |\bm{k} \phi_{\bm{k}}|^{2}
 \left( \Re\left( \frac{\sigma_{s}}{\epsilon_{0}} \right) - \omega_{\ast,n,s} \Im(\chi_{s}) \right).
\label{eq:linear_energy_evolution}
\end{align}
The first and second terms on the right-hand side, respectively, correspond to the energy exchange and injection.

With these preparations, we now examine the stability of the system in terms of energetics following the prescription given in~\cite{Hasegawa_75a}. As in~\cite{RicciRogersDorland_06}, we neglect the electron FLR effect, $k \rho_{\mathrm{e}} \rightarrow 0$, but keep the small but finite ion FLR effect, $k \rho_{\mathrm{i}}\ll 1$. 
 We also assume $\omega/\omega_{B,s}\ll1$ to reduce the plasma dispersion function, $\Xi(z) \approx \imag \sqrt{\pi}-2z$.
 To consider the marginal state, we write $\omega=\omega_{0} + \imag \Im(\omega)$ and assume $\omega_{0} \gg \Im(\omega)$. By Taylor expanding the dispersion relation $\sigma \equiv \sigma_{\mathrm{i}} + \sigma_{\mathrm{e}}=0$ (which is equivalent to $\chi = 0$) around $\omega=\omega_{0}$, the solution must satisfy
\begin{align}
\Im(\sigma(\omega=\omega_{0})) & = - \omega_{0} \epsilon_{0} \Re(\chi(\omega=\omega_{0}))=0, \\
\Im(\omega) & = -\frac{\Re(\sigma(\omega=\omega_{0}))}{\left(-\partial \Im(\sigma)/\partial \omega|_{\omega=\omega_{0}}\right)}.
\label{eq:instability_condition}
\end{align}
The numerator and denominator in \eqref{eq:instability_condition} indicate, respectively, the {\it dissipation} due to wave-particle interactions and the wave energy, $-\partial \Im(\sigma)/\partial \omega \approx \epsilon_{0} \partial \Re(\omega \chi)/\partial \omega$~\cite{LandauLifshitzPitaevskii_84}. Therefore, \eqref{eq:instability_condition} states that the coupling of the negative (positive) dissipation with the positive (negative) energy wave leads to instability---This generalizes the statement that the negative energy is the necessary condition for instability in the systems without wave-particle interactions or other loss mechanisms.

\begin{figure}[htbp]
\centering
\includegraphics[scale=0.5]{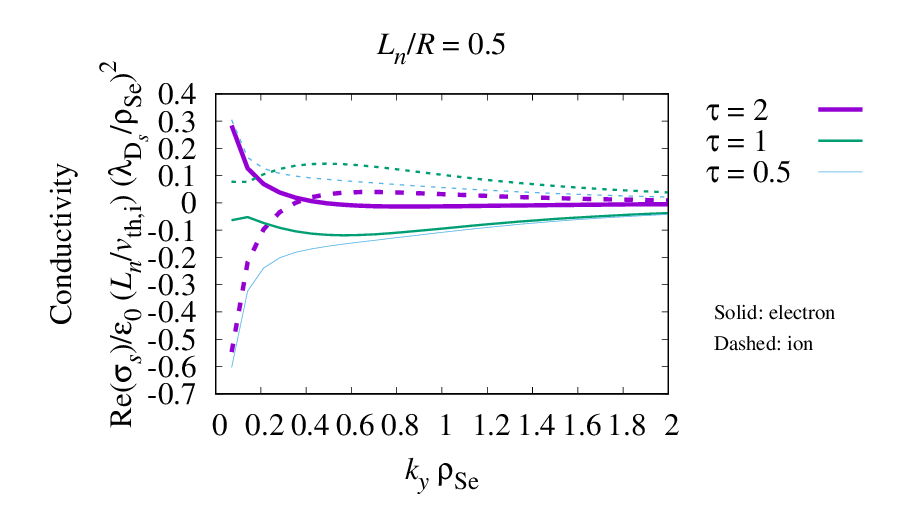}
\caption{\label{fig:sigma}The conductivity calculated from linear gyrokinetic solutions.}
\end{figure}

Care must be taken when approximating the square roots to obtain the solution. According to the solution given in~\cite{RicciRogersDorland_06}, $\omega_{0}>0$ for $\tau/Z<1$ where $\tau\equiv T_{0\mathrm{i}}/T_{0\mathrm{e}}$ is the temperature ratio and $Z\equiv -q_{\mathrm{i}}/q_{\mathrm{e}}$ is the ion charge number (We always set $Z=1$). For positive $\Im(\omega)$, $\sqrt{-\omega/\omega_{B,\mathrm{i}}}\approx\sqrt{(Z/\tau)(\omega_{0}/\omega_{B,\mathrm{e}})}$ and $\sqrt{-\omega/\omega_{B,\mathrm{e}}}\approx-\imag\sqrt{\omega_{0}/\omega_{B,\mathrm{e}}}$. Then, we obtain
\begin{align}
\frac{\omega_{0}}{\omega_{\ast,n,\mathrm{e}}} & = \frac{\tau}{Z} \frac{1}{2\pi(R_{\mathrm{{c}}}/L_{n})^{3}}
\nonumber \\
    & ~~~ \times
    \left[\left(1+\frac{\tau}{Z}\right)\left(\pi\frac{R_{\mathrm{c}}}{2L_{n}}-1\right)-
        \frac{k^{2}\rho_{\mathrm{i}}^{2}}{2} \frac{R_{\mathrm{c}}}{2L_{n}}(\pi-2) \right]^{2},
        \\
\frac{\Im(\omega)}{\omega_{\ast,n,\mathrm{e}}} & = 2 \left(\frac{\tau}{Z}\right)^{3/2}
    \frac{\omega_{0}}{\omega_{\ast,n,\mathrm{e}}},
\end{align}
which is the solution for $\tau/Z\ll1$ of~\cite{RicciRogersDorland_06}.
A positive imaginary part results because
\begin{align}
\Re(\sigma(\omega=\omega_{0})) & = - \omega_{0} \frac{\epsilon_{0}}{k^{2}\lambda_{\mathrm{D},\mathrm{e}}^{2}}
4 \sqrt{\pi} \frac{\omega_{\ast,n,\mathrm{e}}}{\omega_{B,\mathrm{e}}}
\sqrt{\frac{\omega_{0}}{\omega_{B,\mathrm{e}}}} < 0, 
\label{eq:electron_dissipation}
\\
-\left.\frac{\partial \Im(\sigma)}{\partial \omega}\right|_{\omega=\omega_{0}}
    & = 
    \frac{1}{2}\sqrt{\frac{Z}{\tau}}
 \frac{\epsilon_{0}}{k^{2}\lambda_{\mathrm{D},\mathrm{i}}^{2}}
4 \sqrt{\pi} \frac{\omega_{\ast,n,\mathrm{i}}}{\omega_{B,\mathrm{i}}}
 \sqrt{\frac{\omega_{0}}{\omega_{B,\mathrm{e}}}} > 0.
\end{align}
(Note that the electron negative dissipation \eqref{eq:electron_dissipation} also indicates that the electron energy is negative. Under the present assumption, $\omega_{0}\gg\Im(\omega)>0$, the negative dissipation and negative energy are equivalent.)
Therefore, we find that, for $\tau/Z<1$, the wave propagates in the electron direction and the system becomes unstable due to the coupling of positive energy ions and negative dissipation electrons. It is evident from \eqref{eq:linear_energy_evolution} that the species having negative conductivity must extract the energy from the background to grow its energy.
By performing the same analysis for $\tau/Z>1$, we find that the positive energy electrons couple with the negative dissipation ions to lead to instability. According to the direction of wave propagation, the roles played by ions and electrons flip. Figure~\ref{fig:sigma} shows the conductivity $\sigma_{s}$ for $\tau=0.5, 1, 2$ and $L_{n}/R_{\mathrm{c}}=0.5$ obtained from the linear gyrokinetic solutions including full FLR of both ions and electrons. We show $\Re(\sigma_{s})/\epsilon_{0} (L_{n}/v_{\mathrm{th,i}}) (\lambda_{\mathrm{D},s}/\rho_{\mathrm{Se}})^{2}$ against the wave number $k\rho_{\mathrm{Se}}=k\rho_{\mathrm{i}}/\sqrt{2\tau}$. For $\tau<1$, we confirm that the electron conductivity is negative as calculated above. For $\tau>1$, the ion conductivity is negative only for small $k\rho_{\mathrm{Se}}$, but it becomes positive as $k\rho_{\mathrm{Se}}$ gets large, maybe because of the stabilizing influence of the FLR effect. The range of negative ion conductivity broadens as $L_{n}/R_{\mathrm{c}}$ becomes large.
For $\tau=1$, the eigenvalue $\omega$ is pure imaginary, and the assumption that $\omega_{0} \gg \Im(\omega)$ is violated. In this case, it is the electron conductivity that is negative as the numerical solution shows. We expect that the sign of conductivity changes slightly above $\tau=1$.

What are the nonlinear consequences of these instability mechanisms? One may speculate that the sign of the conductivity corresponds to the heating direction, i.e., for $\tau/Z<1$, the negative electron conductivity indicates the positive ion heating, and {\it vice versa}. If that is true, the plasma tends to thermally equilibrate between species to achieve an equal temperature state, namely $\tau$ approaches to the unity. We perform nonlinear gyrokinetic simulations to confirm the speculation.

\begin{figure}[htbp]
\centering
\includegraphics[scale=0.5]{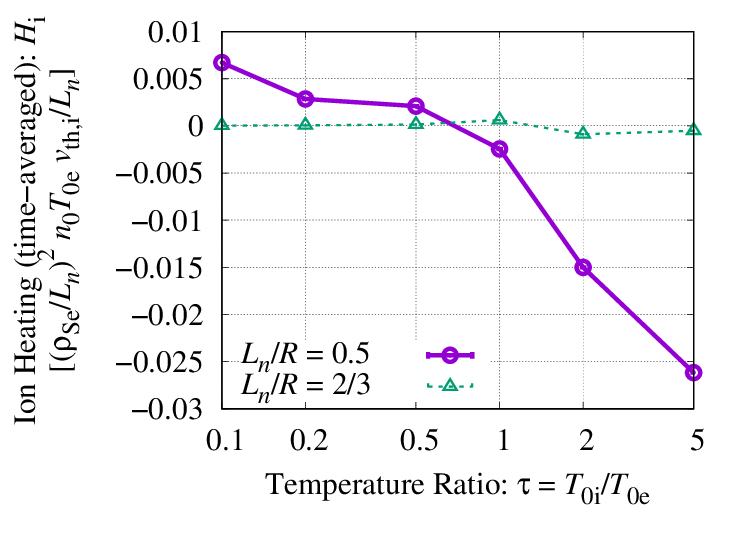}
\caption{\label{fig:heating_ion}The time-averaged ion heating $\overline{H_{\mathrm{i}}}$.}
\end{figure}

Numerical simulations are performed using the {\tt GS2} code~\cite{BarnesDickinsonDorland_22}.
We fix the ion charge $Z=1$, mass ratio $m_{\mathrm{i}}/m_{\mathrm{e}}=1836$, and change the temperature ratio $\tau$ and density gradient $L_{n}/R_{\mathrm{c}}$. The collisionality for ions and electrons are both small, $\nu L_{n}/v_{\mathrm{th,i}}=10^{-2}$. We include the hyper-visicosity~\cite{RicciRogersDorland_06a} to achieve a saturated state. The amplitude is $D_{\mathrm{hyp}}L_{n}/v_{\mathrm{th,i}}=1$ and the cut-off wavenumber is $\approx 0.8 k_{\mathrm{max}}$. (The maximum wavenumber in the simulation is $k_{\mathrm{max}}\rho_{\mathrm{Se}} \approx 2\pi (\rho_{\mathrm{Se}}/L) (N/3)$ where $L$ and $N$ are the box size and grid number.) The simulation domain is $L_{x}/\rho_{\mathrm{Se}}=L_{y}/\rho_{\mathrm{Se}}=40\sqrt{2}\pi$ and the numerical resolution is $(N_{x},N_{y},N_{\lambda},N_{E})=(128,128,8,32)$ where $\lambda$ and $E$ are the velocity space coordinates~\cite{NumataHowesTatsuno_10}. To measure the heating, we use the symmetric form proposed by Candy~\cite{Candy_13},
\begin{align}
H_{s} = \frac{1}{2V} \int q_{s}\left(h_{s} \frac{\partial \langle \phi \rangle_{\bm{R}_{s}}}{\partial t} - \frac{\partial h_{s}}{\partial t} \langle \phi \rangle_{\bm{R}_{s}}\right) \diff \bm{v} \diff \bm{r},
\end{align}
which is equivalent to \eqref{eq:heating} if the long-time average is taken.
Figure~\ref{fig:heating_ion} shows the time-averaged ion heating $\overline{H_{\mathrm{i}}}$ in the saturated states, normalized by $(\rho_{\mathrm{Se}}/L_{n})^{2} n_{0} T_{0\mathrm{e}} L_{n}/v_{\mathrm{th,i}}$. We first discuss the strongly driven case by the steep density gradient, $L_{n}/R_{\mathrm{c}}=0.5$. As expected, the ion heating is positive when $\tau$ is small and decreases as $\tau$ increases. The sign changes around $\tau \approx 1$. The turbulent heating is remarkable only when turbulence is strongly driven. When the density gradient becomes shallower ($L_{n}/R_{\mathrm{c}}=2/3$), the zonal flows self-generated from turbulence almost suppress turbulent fluctuations and resultant heating. (By inspecting the numbers, we still see the positive (negative) ion heating for $\tau\lesssim1$ ($\tau\gtrsim1$).)

We have considered the entropy mode turbulence in the Z pinch configuration. It has been shown that the drift wave propagation direction and responsible species for destabilization change according to the temperature ratio $\tau$. If the ion temperature is larger than that of electrons, ions take the energy from the background to heat electrons, and {\it vice versa}. Therefore, the plasma tries to equalize the temperature even without collisions.
Nonlinear simulations confirm that the preferable temperature ratio is $\tau \approx 1$.
We have examined heating in a strongly turbulent state driven by the steep density gradient. However, if the turbulence is weak, the zonal flows regulate transport and also heating.
To comprehensively understand the self-organized state in the magnetosphere, the transport and meso-scale structure formation, as well as the newly proposed energy exchange mechanism, must be taken into account simultaneously.
We also note that it may be over-simplification to employ the Z pinch as a proxy for the magnetospheric plasma although they share the fundamental instability mechanism. How the dynamics along the magnetic field changes the heating mechanism is open, which we leave for future consideration.
As the final remark, we comment on diagnosing the heating mechanism presented here. The wave-particle interaction occurring in the entropy mode generates a unique structure in velocity space, a signature of the resonance along $\omega=\omega_{\mathrm{D},s}$ as shown in~\cite{RicciRogersDorland_06}.
The instability mechanism of the entropy mode may be directly measured by a spacecraft using the field-particle correlation technique~\cite{KleinHowes_16}.

This work was supported by JSPS KAKENHI Grant Number JP22K03568. Numerical simulations were performed on "Plasma Simulator" (NEC SX-Aurora TSUBASA) of NIFS with the support and under the auspices of the NIFS Collaboration Research
program (NIFS22KISS019).

\providecommand{\noopsort}[1]{}

\end{document}